\documentclass[aps,prd,preprint,eqsecnum,showpacs,showkeys,byrevtex]{revtex4}

\usepackage{amsmath}
\usepackage{bm}
\begin{document}


\title{On the instability of classical dynamics \\ in theories with higher derivatives}



\author{V.~V.~Nesterenko}
\affiliation{Bogoliubov Laboratory of Theoretical Physics, Joint
Institute for Nuclear Research, 141980 Dubna, Russia}
\email{nestr@theor.jinr.ru}


\date{\today}

\begin{abstract}
The development of instability in the dynamics of theories with
higher derivatives is traced in detail in the framework of the
Pais-Uhlenbeck fourth oder oscillator. For this aim the external
friction force is introduced in the model and the relevant
solutions to equations of motion are investigated. As a result,
the physical implication of the energy unboundness from below in
theories under consideration is revealed.

\end{abstract}


\maketitle

\section{Introduction\label{Intr}}
All fundamental physical theories are defined by differential
equations at most of the second order. However, in the course of
searching for new theories the models described by higher-oder
derivative Lagrangians are considered too~\cite{PU,Stelle}. It
is worth mentioning here the gauge theories with higher
derivatives~\cite{gauge1,gauge2}, the gravitation models with
higher oder curvature corrections to the Einstein-Hilbert
action~\cite{BOS,Gr3,Gr4,Gr5,Gr6,Gr7,Gr8,Woodard}, the models of point particles with
Lagrangians depending on curvature and torsion of the world
trajectories~\cite{Nestr,Plyush}, rigid
strings~\cite{NNSH}. The higher derivative theories have
some appealing properties, in particular, the convergence of the
Feynman diagrams is improved~\cite{Thirring}.

 An important peculiarity of higher derivative theories is the
energy unboundness from below  already at the classical level.
Here the energy is defined as the conserved Noether quantity
corresponding to the invariance of the theory under time
translation or, that is the same, as the value of the Ostrogradski
Hamiltonian on the solutions to the equations of
motion~\cite{ChN,NVV1}. It is agreed-upon that the bottomless energy spectrum
in theories with higher derivatives results in their instability.
However the development of this
instability was not traced in detail in literature yet.

 If the theory with higher derivatives is conservative, then the
negative value of its energy, which is preserved in time, does
not formally lead to any contradictions because it is
unobservable simply. In experiment we can detect only the
energy changes. However when such a system experiences
an external action by dissipative forces, the following situation is
feasible. The system will accomplish a positive work against the
external friction forces and, at the same time, the amplitude of internal
motion, which gives a negative contribution to the energy, will
not damp, as usual, but it will increase. Remarkably, the energy
conservation law will formally hold  during this process. Such a
behaviour of a material system is obviously absurd from the physical standpoint, and the
pertinent theory should be discarded without any references on
the instability. Nevertheless it is instructive to trace the
development of this instability.

Since the energy in theories with higher derivatives is not
restricted from below, there are no
reasons to stop the increase of the amplitude of internal motion mentioned above.
Hence a weak external action can lead to a drastical change of
the internal dynamics of the system under consideration. It is
this behaviour of the dynamical system described by higher-order
Lagrangian function that should be kept in mind when saying
about instability of such systems.

It is worth noting that at the quantum level this feature of the theories with higher
derivatives is manifested usually through the appearance of the
states with negative norm~\cite{PU,Heisenberg,ChN,Woodard} and, as a consequence, in
violation of the unitarity. However the physical origin of this drawback is
the bottomless energy spectrum  in  theories at hand.

This paper seeks to trace in detail the development of this
instability considering the Pais-Uhlenbeck fourth oder
oscillator~\cite{PU} which is the simplest theory of this kind.

The layout of the paper is as follows. In Sec.\ II the
Lagrangian and Hamiltonian descriptions of the Pais-Uhlenbeck
fourth order oscillator are presented. In Sec. III the classical
dynamics of this oscillator with allowance for the friction
force is investigated. The development of the instability is traced in detail.
In Conclusion (Sec.\ IV) the obtained results are
summarized and their relation to other studies in this field are noted.

\section{Pais-Uhlenbeck fourth order oscillator: Lagrangian and Hamiltonian descriptions}
The simplest higher derivative theory of a scalar field $\varphi(t,\mathbf{x})$ is defined
by the equation~\cite{PU}
\begin{equation}
\label{2-1}
\left (
\frac{\partial ^2}{\partial t^2}-\Delta +m_1^2
\right )
\left (
\frac{\partial ^2}{\partial t^2}-\Delta +m_2^2
\right )\varphi(t, \mathbf{x})=0\,{,}
\end{equation}
where $\Delta$ is the Laplace operator,  $m_1$ and $m_2$ are the mass parameters.
Upon passing to the momentum space
\begin{equation}
\label{2-2}
\varphi(t,\mathbf{x})=\int e^{i\mathbf{p}\mathbf{ x}} \tilde \varphi
(t,\mathbf{p})\, d^3\mathbf{p}
\end{equation}
equation (\ref{2-1}) acquires the form
\begin{equation}
\label{2-3}
\left (
\frac{d^2}{d t^2} +\omega_1^2
\right )
\left (
\frac{d^2 }{d t^2}+\omega_2^2
\right )\tilde \varphi(t, \mathbf{p})=0\,{,}
\end{equation}
where $\omega _i^2=\mathbf{p}^2+m_i^2, \quad i=1,2$.

When the space has zero dimension, we obtain from (\ref{2-3}) the equation governing
the dynamics of the respective point  mechanical analog of the initial
 field theory (\ref{2-1}),
i.e., the Pais-Uhlenbeck fourth order oscillator
\begin{equation}
\label{2-4}
\frac{d^4 x}{dt^4}+(\omega_1^2+\omega_2^2)\frac{d^2 x}{dt^2}+\omega_1^2\omega_2^2\,x=0{\,},
\end{equation}
where $x\equiv x(t)$ is the coordinate of this oscillator.

The general solution to this equation has the form
\begin{equation}
\label{2-5}
x(t)=x_1(t)+x_2(t),
\end{equation}
where
\begin{equation}
\label{2-6}
x_n(t)=a_n e^{i\omega_n t}+ a^*_ne^{-i\omega_n t}, \quad n=1,2\,{.}
\end{equation}
Here $a_n$ are complex amplitudes, the positive  frequencies $\omega_1$
and $\omega_2$ are assumed to be different and $\omega
_2>\omega_1$. The case of equal frequencies $\omega_1=\omega_2$
requires a special consideration~\cite{Smilga,Mannheim}.

Equation (\ref{2-4}) is the Euler equation
\begin{equation}
\label{2-7}
\frac{\partial L}{\partial x}-\frac{d}{dt}\frac{\partial L}{\partial \dot x}+\frac{d^2}{dt^2}
\frac{\partial L}{\partial \ddot x}=0
\end{equation}
for the Lagrangian
\begin{equation}
\label{2-8}
L=-\frac{1}{2}\, \ddot x^2+\frac{1}{2}\,(\omega_1^2+\omega_2^2)\,\dot x^2-\frac{1}{2}
\,\omega_1^2\omega_2^2\,x^2{.}
\end{equation}

Transition to the Hamiltonian description is accomplished by
making use of the Ostrogradski method~\cite{Ostr}.
Upon introducing the canonical variables
\begin{eqnarray}
\label{2-9}
q_1&=&x, \nonumber  \\
q_2&=& \dot x, \nonumber \\
p_1&=& \frac{\partial L}{\partial \dot x} -\frac{d}{dt}\frac{\partial L}{\partial \ddot x}=(\omega_1^2
+\omega_2^2)\,\dot x+\ddot x{,} \nonumber \\
p_2&=& \frac{\partial L}{\partial \ddot x} =-\ddot x
\end{eqnarray}
the Ostrogradski Hamiltonian is constructed in the usual way
\begin{equation}
\label{2-10}
H=p_1\dot q_1+p_2\dot q_2-L=\frac{p_1^2}{2(\omega_1^2+\omega_2^2)}+\frac{1}{2}\,
\omega_1^2\omega_2^2 q_1^2 -\frac{1}{2}p_2^2-\frac{1}{2}(\omega_1^2+\omega_2^2)\left (
q_2-\frac{p_1}{\omega_1^2+\omega _2^2}
\right )^2{.}
\end{equation}
Obviously, this Hamiltonian is not positive definite.

The Hamiltonian equations of motion have the form
\begin{eqnarray}
\dot q_1&=&\frac{\partial H}{\partial p_1}=q_2\,{,} \nonumber \\
\dot q_2&=&\frac{\partial H}{\partial p_2}=-p_2\,{,} \nonumber \\
\dot p_1&=&-\frac{\partial H}{\partial q_1}=-\omega_1^2\omega_2^2q_1\,{,} \nonumber \\
\dot p_2&=&-\frac{\partial H}{\partial q_2}=(\omega_1^2+\omega_2^2)\,q_2-p_1\,{.}
\label{2-11}
\end{eqnarray}
Twice differentiating  the second equation in (\ref{2-11}) and making use of the rest Hamiltonian
 equations one deduces the Lagrangian equation (\ref{2-4})
\begin{eqnarray}
\label{2-12}
\dddot q_2\equiv \frac{d^4 x}{dt^4}=\frac{d}{dt} (-\dot p_2)
&=&-\frac{d}{dt}\left [
(\omega_1^2+\omega_2^2)\,q_2-p_1
\right ]=-(\omega_1^2+\omega_2^2)\,\dot q_2+\dot p_1 \nonumber \\
&=&-(\omega_1^2+\omega_2^2)\ddot x-
\omega_1^2\omega_2^2\,x\,{.}
\end{eqnarray}

The Ostrogradski Hamiltonian (\ref{2-12}) can be rewritten in terms of the
Lagrangian variables  $x,\, \dot x, \,\ddot x,\, \dddot x$:
\begin{equation}
\label{2-13}
E=\frac{1}{2}\, \omega^2_1\omega _2^2\, x^2+\frac{1}{2}\,
(\omega _1^2+\omega _2^2)\,\dot x^2 -
\frac{1}{2} \,\ddot x^2+\dot x\,\dddot x\,{.}
\end{equation}
Substituting in this equation the solutions $x_1(t)$ and
$x_2(t)$ from (\ref{2-6}) we obtain the values of the energy for
oscillations with frequencies $\omega _1$ and $\omega_2$,
respectively
\begin{eqnarray}
E_1=(a_1a^*_1+a_1a^*_1)\,\omega _1^2\, (\omega _2^2-\omega _1^2)\,{,}
\label{2-14}\\
E_2=(a_2a^*_2+a_2a^*_2)\,\omega _2^2\, (\omega _1^2-\omega _2^2)\,{.}
\label{2-15}
\end{eqnarray}
It is obvious that $E_1$ and $E_2$ have the opposite signs. If
the inequality $\omega _2>\omega_1$ holds, as we have assumed
above, then the contribution to the Ostrogradski energy of the
oscillations with the frequency $\omega _1$ is positive while the
same with the frequency $\omega _2$ is negative. The
energy (\ref{2-13}) is an integral of motion, and the fact that
it can acquire negative values does not lead to any physical
contradictions.  A different situation arises when the system under consideration
experiences external action.

\section{Pais-Uhlenbeck oscillator with damping}
Let us introduce the friction force into the equation of motion (\ref{2-4})
\begin{equation}
\label{3-1}
\frac{d^4 x}{dt^4}+(\omega_1^2+\omega_2^2)\,\frac{d^2 x}{dt^2}+\omega_1^2\omega_2^2 \,x+\gamma
\frac{dx}{dt}=0\,{,}
\end{equation}
where $\gamma$ is a positive constant (friction coefficient). As before the time
dependence of the
solutions to Eq.\ (\ref{3-1}) is described by the factor $e^{i\omega t}$, where the
frequency $\omega$ is the root of the equation
\begin{equation}
\label{3-2}
\omega ^4 - (\omega _1^2+\omega _2^2)\,\omega ^2 +\omega _1^2\omega _2^2 +2i\gamma \omega
=0\,{.}
\end{equation}

In order to simplify the  calculations we assume that the damping is weak, so the
perturbation theory is applicable. Substituting  in Eq.\ (\ref{3-2})
\begin{equation}
\label{3-3}
\omega=\omega_n+i\Delta \omega_n,\quad n=1,2\,{,}
\end{equation}
we arrive at the equation for $\Delta \omega_n$. In the approximation  linear in $\gamma$
it reads
\begin{equation}
\label{3-4}
2\,\omega^2_n\Delta \omega_n- (\omega _1^2+\omega _2^2)\,\Delta \omega_n +\gamma=0,\quad
n=1,2\,{.}
\end{equation}
Thus, the frequency shifts for $\omega_1$  and $\omega_2$ prove
to be real and equal in the absolute value but they have the  opposite signs
\begin{eqnarray}
\Delta \omega _1=\frac{\gamma }{\omega _2^2-\omega _1^2} >0\,{,}
\label{3-5}\\
\Delta \omega _2=\frac{\gamma }{\omega _1^2-\omega _2^2} <0\,{,}
\label{3-6}
\end{eqnarray}
because we have assumed that $\omega _2>\omega_1$.
As a result, the solution
\begin{equation}
\label{3-7}
x_1(t) \sim e^{i\omega_1 t-\Delta \omega_1 t}
\end{equation}
exponentially decreases in time, while the solution
\begin{equation}
\label{3-8}
x_2(t) \sim e^{i\omega_2 t+|\Delta \omega_2| t}
\end{equation}
exponentially increases in time. The damping behaviour of the solution
$x_1(t)$ is physically motivated, namely, due to the external
friction force the amplitudes $a_1$ and $a_1^*$ in Eq.\
(\ref{2-14}) decrease in time and the energy $E_1$ also
decreases being positive. Unlike this, the amplitudes $a_2$ and
$a_2^*$ in Eq.\ (\ref{2-15}) exponentially grow up and there are
no reasons to stop this process, since the energy $E_2$, being
negative, decreases without bound. Obviously such a dynamical
behaviour is not acceptable from the physical stand point.
Remarkably,  for the both solutions $x_1(t)$ and $x_2(t)$ the
energy is conserved however in the latter  case this holds only
formally.

Now we consider the Pais-Uhlenbeck oscillator which  experiences
an arbitrary force $f(t)$:
\begin{equation}
\label{3-9}
\frac{d^4 x}{dt^4}+(\omega_1^2+\omega_2^2)\frac{d^2 x}{dt^2}+\omega_1^2\omega_2^2
\,x=f(t)\,{.}
\end{equation}
The general solution to this equation
can be expressed in terms of the relevant Green function
\begin{equation}
\label{3-10}
x(t)= \int\limits_{-\infty}^{\infty}G(t-t')\,f(t')\,dt'
= \int\limits_{-\infty}^{\infty}\bar G(\omega)\,
\bar f(\omega)\, d\omega\,{,}
\end{equation}
where
\begin{equation}
\label{3-11}
G(t)= \frac{1}{\sqrt{2\pi}} \int\limits_{-\infty}^{\infty}e^{i\omega t}
\bar G(\omega)\,d\omega, \quad
f(t)= \frac{1}{\sqrt{2\pi}} \int\limits_{-\infty}^{\infty}e^{i\omega t}
\bar f(\omega)\,d\omega\,{.}
\end{equation}
From Eq.\ (\ref{3-9}) we deduce in a straightforward way
\begin{equation}\label{3-12}
\bar G(\omega)= \frac{1}{\omega_1^2-\omega_2^2}\left (
\frac{1}{\omega^2-\omega_2^2}-\frac{1}{\omega^2-\omega_1^2}
\right){.}
\end{equation}
Now we see from Eqs.\ (\ref{3-10}) and (\ref{3-12}), that the forces with spectral densities localized around
 $\omega_1^2$ and $\omega _2^2$ give  rise to displacement $y$ of  opposite signs. Obviously,
it implies that one of these displacements is unphysical.

\section{Conclusion}
By making use of a simple and clear example, tractable
analytically,  we have shown that the instability of the
theories with higher derivatives implies the following.
Introduction of arbitrary weak interaction of such systems
with surroundings  results, for
sure, in  appearance of unbounded rising solutions. Being negative, the
Ostrogradski energy for
these solutions grows up in absolute value without bound
and there are no reasons to terminate this
process.

It is worth noting that introduction of external {\it nonpotential} forces
is crucial in our consideration. Indeed, in Ref.\  \cite{LMP}
an additional potential term was introduced in undamped
Pais-Uhlenbeck oscillator.
It was shown there that indefiniteness of the energy does not forbid the stability
in this case.

  Our inference does not imply at all that any theory with
higher derivatives is unphysical one at the classical level
already. Indeed, let us consider, as an example, a dynamical
theory, which is described by two ordinary differential
equations of the second order. Let the dynamics of this system
be stable (a conservative system with energy bounded from
below). In the general  case, these equations can be transformed
into one differential equation of the fourth order. It maybe
that the Ostrogradski energy for this equation proves to be
unbounded from below. But it does not imply that the initial
system of two equations is unstable with respect to external
action. The point is, the Ostrogradski energy for the one  forth
order equation does not coincides, in the general case, with the
usual energy for two initial differential equations of the second order. The
examples of such systems are discussed now in the literature.
First of all, it is worth noting here the Timoshenko beam theory
which describes the transverse vibrations of an elastic  beam or rod with
allowance for flexure bending and transverse shear
deformation~\cite{NVV,book,Stephen1,Stephen2,ChN}.

\begin{acknowledgments}
The discussion of Ostrogradski's instability and related topics
with Richard Woodard, Neil Stephen, and Kwok Tung Chan   is
acknowledge with a gratitude.  This work was supported  by the
Russian Foundation for Basic Research (Grant No.\ 06-01-00120).
\end{acknowledgments}

\end{document}